\documentclass[prd,a4paper,aps,epsfig,showpacs,twocolumn,floats]{revtex4}

\newcommand{\be}{\begin{equation}}
\newcommand{\ee}{\end{equation}}
\newcommand{\bea}{\begin{eqnarray}}
\newcommand{\eea}{\end{eqnarray}}
\newcommand{\mn}{{\mu\nu}}

\begin{document}
\title{Bimetric varying speed of light theories and
primordial fluctuations}
\author{Jo\~ao  Magueijo}
\affiliation{
Theoretical Physics Group, Imperial College, London, SW7 2BZ}
\date{\today}
\begin{abstract}
{We exhibit a varying speed of light (VSL) theory that implements
the recently proposed decaying speed of sound mechanism for
generating density fluctuations. We avail ourselves of bimetric
VSL theories, where the speed of gravity differs from that of
light. We first show that a DBI-like type of
$K$-essence model has the necessary speed of sound profile to produce
(near) scale-invariant fluctuations. We then examine the map
between bimetric and $K$-essence models: typically the bi-scalar
connecting the two metrics is a $K$-essence field in one of them.
Remarkably, the DBI model is found to perturbatively represent the
minimal bimetric model, where the bi-scalar is Klein-Gordon in the
matter frame. But the full non-perturbative bimetric structure is
even simpler: the bi-scalar dynamics should be simply driven by a
cosmological constant in the matter frame, balanced by an opposite
cosmological constant in the gravity frame. Thus the problem of
structure formation receives an elegant and universal solution
within bimetric VSL theories, which are known to also solve the
flatness and entropy problems and evade a plethora of causality
concerns. }
\end{abstract}
\pacs{0000000}
\maketitle

\bigskip

\section{Introduction}

In a recent letter~\cite{csdot}  we proposed a mechanism for producing
scale-invariant density fluctuations of appropriate amplitude
based on a decaying speed of sound $c_s$.
We emphasized the mechanism's generality and how it could
be implemented using a variety of methods. The examples of
$K$-essence~\cite{kappa,garriga}
and varying speed of light (VSL)~\cite{vsl,jm,am}
were given, and other possibilities,
such as non-adiabatic hydrodynamical matter, were suggested.
But we stressed the generality of the proposal and deliberately chose
not to marry it to any specific model.
It was nevertheless noted that important issues, such as the causality concerns
raised by the presence of superluminal signals or the status of the other
Big Bang problems (such as the flatness or entropy problems)
did require the context of a specific model. In this paper we
investigate specific models, with the focus heavily trained on varying
speed of light theories.

VSL theories may raise disturbing fundamental issues~\cite{caus,ellis,reply},
at worst breaking every symmetry one can think of. But this
need not be the case; indeed they may be simply regarded as
non-trivial realizations of the local
Lorentz group, without introducing preferred frames and other
``oddities''. Two implementations of this idea stand out:
bimetric VSL theories and deformed
special relativity. In the
former the Lorentz group is realized by separate metrics for
matter and for gravity, leading to different speeds for photons
(and other massless ``matter'' particles) and gravitons~\cite{bim,bim1}.
In the latter the dispersion relations are deformed from their usual
quadratic form, so that the
(group) speed of all massless particles becomes energy dependent~\cite{dsr}.
To prevent the introduction of a preferred frame one then chooses a suitable
non-linear representation of the Lorentz group~\cite{ljprl}.
In two companion papers we examine the connection between the varying
speed of sound mechanism of~\cite{csdot} and these two types of VSL theories,
concentrating on the bimetric approach here (see~\cite{amend,piao,next}
for connections with the latter).

Bimetric theories are perhaps the most straightforward
arena for realizing the varying $c_s$ mechanism for
producing scale-invariance.  In~\cite{csdot} we exhibited an
admittedly rather cumbersome $K$-essence theory with the required
$c_s$ profile. Upon closer inspection, however, this
model is found to be a limiting case of the Dirac-Born-Infeld (DBI)
model (with the sign of one of its
constants reversed with respect to what is usual, so we should perhaps
call it ``anti-DBI''), as we prove in this
paper. Furthermore
it has been shown that $K$-essence models are relieved from causality
paradoxes if interpreted as bimetric models~\cite{caus,kbim}.
In this paper we perform this exercise to discover that the DBI
model {\it universally} associated with scale-invariance is precisely
the minimal bimetric model, where the field linking the two metrics
is provided with the simplest possible dynamics.

The plan of this paper is as follows.
We first review bimetric theories that fall under the VSL
remit (Section~\ref{SBim}), stressing a couple of details
of technical relevance for this paper.
 We then recall 
the varying $c_s$ mechanism of~\cite{csdot}, highlighting  the requirements
 for a viable theory of structure formation
(Section~\ref{csdotmech}). We examine the rather wooden
$K$-essence model previously proposed as ``proof of concept''
and show that this is in fact the limiting case of a DBI-like model
(Section~\ref{ksuccess}). With this in mind we then consider the perturbative
map between $K$-essence models and
bimetric theories (Section~\ref{mapping}), showing that
the DBI model considered maps onto a simple, minimal bimetric
model. A few non-perturbative features of bimetric theories are
not captured by this map; but after identifying them we're able
in Section~\ref{full} to write down the full non-perturbative
bi-metric structure behind this model, which is simply
a cosmological constant with respect
to the matter metric.
We conclude with a discussion of other possibilities motivated
by this encouraging result.

Throughout this paper we shall use a metric with signature $( + -
- -)$; this is for consistency with the majority of the literature
quoted, but the reader is warned that some translating was required
with regards to the DBI literature. We shall use the reduced
Planck mass $M_{Pl}=1/(8\pi G)$ and set the present speed of
light to 1.

\section{Bimetric VSL theories }\label{SBim}
Bimetric theories have been proposed as VSL
theories~\cite{bim,bim1}, solving the horizon, flatness, and dark
matter problems. In the model proposed by Clayton and Moffat, for
example, there are two metrics: an ``Einstein'' metric
$g_{\mu\nu}$ (which is used to construct the Einstein-Hilbert
action and defines the ``Einstein'' frame, also called
 ``VSL frame'' in this context);
and a ``matter'' metric ${\hat g}_{\mu\nu}$ (used to
construct the matter action via minimal coupling, and defining the ``matter''
or ``constant $c$'' frame). The left hand
side of Einstein's equations is unmodified with regards to metric
$g_{\mu\nu}$; but matter is minimally coupled to  ${\hat
g}_{\mu\nu}$.

This set up is not unusual and has been used, for instance,
in Brans-Dicke varying $G$ theories~\cite{bdicke}, where the two
metrics define the Einstein and Jordan frames. What is
peculiar to VSL bimetric theories is that the two metrics are
related by a disformal transformation, so that the light-cones
associated with them don't coincide and the speed of
``light'' differs from that of ``gravity''. TeVeS theories
(an alternative to dark matter) can be seen as bimetric
VSL theories~\cite{teves}.

The disformal transformation between the two metrics can be ruled
by a choice of dynamics. In the simplest case one invokes a
``bi-scalar'' field $\phi$ and sets: \be\label{gmnhat} {\hat
g}_{\mu\nu}=g_\mn+B\partial_\mu\phi\partial_\nu\phi\; . \ee Here
$B$ is chosen to have dimensions of $M^{-4}$, so that $\phi$ has
dimensions of $M$. In the most general case $B$ could be any
function of $\phi$ and $X=\frac{1}{2}\partial_\mu\phi \partial^\mu
\phi$, but in the minimal theory it is a constant. The action
breaks into $S=S_g+S_m+S_\phi$ with: \bea
S_g&=& \frac{M_{Pl}^2}{2}\int d^4 x {\sqrt {-g}}\, R[g_{\mu\nu}]\nonumber\\
S_m&=&\int d^4 x {\sqrt {-{\hat g}}}\, {\cal L}_m[{\hat
g}_{\mu\nu},\Phi_{Matt}]  \eea and $S_\phi$ determining the bi-scalar dynamics.
In the original formulation, due to Clayton
and Moffat, $\phi$ has a Klein-Gordon Lagrangian
in the Einstein ($g_{\mu\nu}$) frame, that is
\be\label{phikg1} S_\phi=\int d^4 x {\sqrt {-g}}
\,{\left(\frac{1}{2}g^{\mu\nu}\partial_\mu\phi
\partial_\nu \phi -V\right)}\; .\ee
But we could also make it
Klein-Gordon in the matter (${\hat g}_{\mu\nu}$)
frame: \be \label{phikg2}S_\phi=\int
d^4 x {\sqrt {-{\hat g}}} \,{\left(\frac{1}{2}{\hat
g}^{\mu\nu}\partial_\mu\phi
\partial_\nu \phi -V\right)}\ee or even consider hybrids, for which
kinetic and potential terms live on different frames, e.g. \be
\label{phikg3}S_\phi=\int d^4 x {\sqrt {-{\hat g}}}
\,\frac{1}{2}{\hat g}^{\mu\nu}\partial_\mu\phi
\partial_\nu \phi - \int d^4 x {\sqrt {-g}}
\, V \; .\ee The bi-scalar, after all, connects the two frames,
so any of these dynamics is possible and can be considered minimal
if $B$ is constant. A subtlety, to be unveiled in Section~\ref{full},
reveals an even more ``minimal'' dynamics.

We stress two little known peculiarities of these theories, which
will be of technical relevance later. First the Einstein equations
remain unmodified in the Einstein frame ($G_{\mu\nu}=
-T_{\mu\nu}/M_{Pl}^2$) only if we define the stress energy tensor as \be
T^{\mu\nu}=\frac{2}{\sqrt{-g}}\frac{\delta}{\delta
g_{\mu\nu}}(S_M+ S_\phi) \ee i.e. if we map the matter (and
bi-scalar, if applicable) Lagrangian into the Einstein frame
before computing $T_{\mu\nu}$. The Bianchi identities are then
promptly satisfied. But if we use the matter frame definition:
 \be {\hat
T}^{\mu\nu}=\frac{2}{\sqrt{-{\hat g}}}\frac{\delta}{\delta {\hat
g}_{\mu\nu}}(S_M+ S_\phi) \ee (say, for theory (\ref{phikg2}))
then we have instead \be G^{\mu\nu}=-\frac{1}{M_{Pl}^2} {\sqrt{\frac{\hat
g}{g}}}{\hat T}^{\mu\nu} \ee and consistency with the Bianchi
identities is more intricate to prove~\cite{bim}. Cosmological
perturbation calculations are most easily performed in the Einstein frame, so
it is important to bear this in mind.

A second peculiarity is that even though matter is minimally
coupled to ${\hat g}_{\mu\nu}$ {\it for both Lagrangian and
equations of motion} the same cannot be said of the bi-scalar. If
the bi-scalar has a Klein-Gordon Lagrangian (either in the
Einstein or matter frames) its field equations won't  be
Klein-Gordon, but receive a correction. This is because, upon
performing variations in $S$, one finds \be \delta {\hat
g}_{\mu\nu}=\delta g_{\mu\nu}+ 4B\partial_{(\mu} \phi
\partial_{\nu )}\delta\phi \ee so that  a term related to the
stress-energy tensor appears in the field equation of $\phi$. If we
take (\ref{phikg1}) this becomes \be\label{modkg1}
\nabla^2\phi+V'(\phi)+{\sqrt{\frac{\hat g}{g}}}B{\hat
T}_M^{\mu\nu} {\hat \nabla}_\mu{\hat \nabla}_\nu \phi=0\; . \ee
This led~\cite{bim} to the introduction of a ``third'', derived
metric, associated with the Klein-Gordon propagation of $\phi$
(see their papers for details). If we take (\ref{phikg2}) 
the field equation becomes: \be\label{modkg2} {\hat
\nabla}^2\phi+V'(\phi)+ B{\hat T}^{\mu\nu} {\hat \nabla}_\mu{\hat
\nabla}_\nu \phi=0 \ee so that the ``third metric'' providing the
field with a Klein-Gordon type of equation: \be
\label{pseudokg}{\overline g}^{\mu\nu}{\hat \nabla}_\mu{\hat
\nabla}_\nu\,\phi +V'(\phi)=0\;
 \ee
is
\be {\overline g}^{\mu\nu}={\hat g}^{\mn}+B{\hat T}^\mn \; .\ee
This is not a proper space-time structure;
 for instance the covariant derivative
$\hat \nabla_\mu$ used in (\ref{pseudokg}) is still defined with
respect to metric ${\hat g}_\mn$. But it renders the field equations
in a Klein-Gordon format, something that will be of use 
later~\footnote{Note that the equation and third metric
for (\ref{phikg3}) have a similar form.}.

It is a notable fact that (in the absence of  matter) the
Lagrangian that does produce a Klein-Gordon equation in the ${\hat
g}_{\mu\nu}$ frame is a pure cosmological constant.  This will be
explained
further in Section~\ref{full}, but it should already be obvious from these
expressions.

\section{The decaying speed of sound mechanism}\label{csdotmech}
We now review the mechanism for generating scale-invariant fluctuations
proposed in~\cite{csdot}. The central result is that for any (constant)
equation of state $w=p/\rho$ scale-invariance follows from a sound speed
$c_s\propto \rho$, if the fluctuations originate
from a vacuum state defined inside the (sound) horizon,
according to the standard prescription.


Let the speed of sound
diverge with conformal time according to $c_s\propto \eta^{-\alpha}$
(with $\alpha>0$; note that $\eta$ is positive and increases from
zero). Whether we employ a hydrodynamical or a scalar field
description, the density
fluctuations are described by a modified harmonic oscillator
equation. This can be written in terms of a variable $v$ related
to the curvature perturbation $\zeta$ by $\zeta=-v/z$, with $z
\propto \frac{a}{c_s}$. The equation for $v$
is~\cite{mukh,lidsey,garriga}: \be\label{veq} v''+\left[c_s^2 k^2
-\frac{z''}{z}\right]v=0\; . \ee
As with inflation this
equation can be solved with Bessel functions,
with solutions: \be\label{vbess}
v=\sqrt{\beta\eta}(AJ_\nu(\beta c_sk\eta)+BJ_{-\nu}(\beta c_s
k\eta))\; . \ee The order $\nu$ is given by \be
\nu=\beta{\left(\alpha-\frac{3(1-w)}{2(1+3w)}\right)} \ee
and $\beta=1/(\alpha-1)>0$.
The boundary conditions, inside the (sound) horizon ($kc_s\eta\gg1$)
can be found from the WKB solution:
\be\label{bc1} v\sim\frac{e^{ik\int c_s d\eta}}{\sqrt{c_s k}}\sim
\frac{e^{-i \beta c_s k \eta}}{\sqrt{c_s k}} \; .\ee
These are satisfied if
the integration constants $A$
and $B$ in (\ref{vbess}) are $k$-independent numbers of order 1.
The spectrum left outside the horizon can now be found.
Since $c_s\eta$ is a
decreasing function of time, the negative order solution is the
growing mode, so that asymptotically we have: \be\label{vout}
v\sim \frac{\sqrt{\beta \eta}}{(c_s k \eta)^{\nu}}\; . \ee
Scale-invariance of the curvature fluctuation
($k^3\zeta^2=const$) therefore requires $\nu=3/2$, i.e.
\be\label{alpha0} \alpha=\alpha_0=6\frac{1+w}{1+3w}\;  \ee
and if we rewrite $c_s$ in terms of the density
$\rho$ we conclude that this implies $c_s\propto \rho$  for all
$w$. The spectrum can also be made red or blue depending on
whether $\alpha<\alpha_0$ or $\alpha>\alpha_0$, specifically
\be\label{nvac} n_S-1=\beta (\alpha-\alpha_0)\; . \ee
Finally the fluctuations' amplitude can be
worked out from formula (see~\cite{garriga}):
\be\label{ampcsrho} k^3\zeta^2\sim
\frac{(5+3w)^2}{1+w}\frac{\rho}{M^4_{Pl}c_s}\; , \ee
giving further support for $c_s\propto \rho$ as the root of
scale-invariance.
At low densities $c_s$ must be constant so scale-invariance must
effectively require $c_s=c_0(1+\rho/\rho_\star)$, where $c_s\approx c_0$ at
low-energy and $\rho_\star$ is the density that triggers its
divergence. Thus the normalization is set by the ratio of this
scale to the Planck scale, viz:
\be\label{amp} k^3\zeta^2\sim
\frac{(5+3w)^2}{1+w}\frac{\rho_\star}{M^4_{Pl}}\sim 10^{-10}\; .
\ee
For the rest of this paper we shall use bimetric VSL theories
to implement this mechanism, but for the sake of clarity
it is important to dissociate our efforts from those
previously made~\cite{mofstruc}.

In~\cite{mofstruc} a structure formation model was based on a
bi-scalar $\phi$ which is Klein-Gordon in the
Einstein frame, i.e. satisfies (\ref{phikg1}) with $V=0$.
Therefore its speed of sound in the Einstein frame (where the fluctuations'
calculation is done) is fixed, $c_s=1$. The fluctuations in $\phi$
follow Eqn. (\ref{veq}) with $c_s=1$ and decelerated expansion,
and so their modes start off frozen-in (outside the horizon),
oscillating only in the current epoch, when they enter the horizon.
The model, thus,  cannot rely on conventional methods for setting the initial
conditions, i.e. by considering (vacuum quantum) fluctuations
inside the (sound) horizon and then following them as they leave the
horizon and freeze-out. Quite the opposite happens:
the $\phi$ modes do not {\it re}-enter the horizon nowadays because they were
never inside the horizon before. For the purpose of structure formation
the horizon problem for $\phi$ has not been solved.
Thus the need to appeal to ``non-conventional'' initial conditions
for super-horizon modes, as considered in~\cite{hollands}.

Undoubtedly the horizon problem is solved for {\it matter} in these models.
But matter is subdominant and irrelevant for structure formation,
as recognized in~\cite{mofstruc}. It may be that interactions between the bi-scalar and
matter causally connect super-horizon modes for the bi-scalar but
these interactions are assumed to be sub-dominant (indeed matter is set to zero
in~\cite{mofstruc}). The main interaction of the bi-scalar is
with gravity (according to (\ref{veq}) with $c_s=1$), and
fluctuation modes in the bi-scalar start frozen-in, not
oscillating, stressing the presence of a ``horizon problem''.

With regards to ``non-conventionality'', we do not feel qualified
to throw the first stone. But we do believe it
unnecessary to invoke new methods for setting initial conditions
to obtain scale-invariance in bimetric models, as
we demonstrate in this paper.

\section{The (anti-)DBI model as a realization of scale-invariance}\label{ksuccess}
In~\cite{csdot} we already exhibited a proof of concept realization
of the scenario just reviewed. Here we  show that this realization
is the limiting case of a Dirac-Born-Infeld theory, with
a crucial change in the sign of one constant with respect to what is
usual in the literature.

The model in~\cite{csdot} is based on $K$-essence, a scalar field
theory with non-trivial kinetic
terms~\cite{kappa,kappa-ruth,vikman,bean}. The Lagrangian has the
form \be {\cal L}=K(X)-V(\phi) \ee (with
$X=\frac{1}{2}\partial_\mu\phi\partial^{\mu}\phi$),
where $K$ can be any function of $X$. Computing  the
stress energy tensor we find the pressure and density:
\bea
p&=&K-V\\
\rho&=&2XK_{,X}-K+V\; \eea
whereas the speed of sound is found to be:
\be\label{csK} c_s^2=\frac{K_{,X}}{K_{,X}+2XK_{,XX}}\; . \ee
The cuscuton model~\cite{cusc1,cusc2} is a good starting point
for our construction. It is defined as a model with $K\propto
\sqrt {X}$. It has a number of peculiarities: its speed of sound
is infinite (since the denominator of (\ref{csK}) vanishes);
its kinetic term $K$ doesn't contribute to $\rho$
but only to $p$, so that its energy density is fully
due to the potential $V$ if present. For a general potential the field
can only take on discrete values if living on its own, but if other
matter is present the cuscaton locks on to its dynamics. An interesting
exception is a cuscuton with a suitably chosen mass term, which 
exhibits scaling solutions even without background matter~\footnote{
If there is background matter the cuscuton with a mass potential always
scales with it. In the absence of matter by scaling we mean a solution
where the density goes like $1/t^2$.}. Specifically,
\be {\cal L}_0=\mu^2\sqrt {|X|}-\frac{1}{2}m^2\phi^2
\ee subject to:
\be
\mu=\frac{2}{\sqrt{3}}\sqrt{m M_{Pl}}\; , \ee
leads to solutions with $\rho\propto 1/t^2$ and $\phi=A/t$.
These scaling solutions have a stable
equation of state:
\be\label{eqstat}
w=\sqrt{\frac{8}{3}}\frac{M_{Pl}}{m A}\; .
\ee
but the model differs from the structure formation
requirements in that $c_s=\infty$.
In order to implement  $c_s\propto \rho$ a
term of form ${\cal L}_1\propto X^n$
with $n< 1/2$ should be added to $K$. This doesn't affect
the homogeneous solution and its equation of state at high
energies, but $c_s$ is no longer infinite. Using Eqn.~(\ref{csK}) we find \be
c_s^2\propto X^{\frac{1}{2}-n}\propto \rho^{1-2n}\; . \ee so that
for $n=-1/2$ we have $c_s\propto \rho$, as required for
scale-invariance.

This argument may look contrived, but we repeated it here because
it is ``constructive'' with regards to what is required from the
point of view of producing fluctuations. But we now show that
it can be directly motivated from the DBI action. As abundantly
pointed out in the literature~\cite{dbi1,dbi2,dbi3,dbi4} this
model can be associated with ``stringy physics'' (an association
which may or may not be relevant for this paper). DBI inflation
has been studied in recent papers, where the model is tweaked so
as to {\it reduce} the speed of sound during inflation. However
the model could also be used to implement the varying speed of
sound mechanism, dispensing with inflation altogether. The DBI
model for a scalar field is based on the action: \be\label{dbiS} {\cal
L}=-\frac{1}{f(\phi)}\sqrt{1-2f(\phi) X}+\frac{1}{f(\phi)}
-V(\phi)\; , \ee where we note that we're using signature $+---$
(in agreement with all other literature used in this paper), as
opposed to what has been used in the literature in this field. A
positive $f$ has the effect of limiting the curvature and reducing
the speed of sound. We choose $f=-C$ where $C$ is a positive
constant, so as consider the minimal model with the opposite
feature: the speed of sound increases at high energies, without
any curvature capping. For $X\ll C^{-1}$ the theory has the
appropriate limit ${\cal L}\approx X-V$, so there's no question of
introducing ghosts due to the signal of $f$. Usually $X$ cannot
grow above a given value ($X=1/2f$), but now it makes sense to
explore the limit $X\gg C^{-1}$ for which we obtain: \be {\cal L}=
\mu^2{\sqrt{X}}+\frac{\mu^6}{4\sqrt{X}} -V(\phi)\; . \ee which is
precisely the ad-hoc $K$-essence model we constructed above (with
$\mu^2=\sqrt{2/C}$), subject to a constraint upon the coefficients
of the terms in ${\sqrt{X}}$ and $1/{\sqrt{X}}$. This finally lets
us determine the value of $\rho_\star$, so important for the
normalization of the spectrum, in terms of Lagrangian parameters:
\be \rho_\star=\frac{3}{8}A\frac{\mu^6}{M_{Pl}^2}=\frac{1}{\sqrt
2} \frac{\mu^4}{w}\; . \ee In the latter identity we used
(\ref{eqstat}) to eliminate $A$ in terms of $w$. The normalization
implied by (\ref{amp}) therefore requires that: \be\label{amp1}
\frac{(5+3w)^2}{\sqrt{2}w(1+w)}\frac{\mu^4}{M_{Pl}^4}\sim 10^{-10}
\ee relating field theory parameters and observables.

Is there anything wrong with choosing the opposite sign for the 
DBI coupling constant? As explained above, there are no ghosts
in the theory, provided a corresponding change in signs is introduced
elsewhere in the Lagrangian (so that the low energy limit still comes
out right). But the theory certainly {\it doesn't come from 
string/brane theory}, since the signature of the space fixes the 
sign of $f$ to be positive (at least using a minimal adaptation of the 
argument). This is not a reason to discard it. The only problem
with the theory seems to be exactly the feature we want to implement:
a superluminal speed of sound. This is a more general problem with
some $K$-essence models, which we now proceed to analyse.

\section{The mapping between $K$-essence and bimetric models}
\label{mapping}
We now  relate the two leitmotifs in this paper: bimetric
VSL theories and the varying speed of sound mechanism as implemented
by the DBI model. $K$-essence models can be seen as theories with an
emergent second metric~\cite{kbim}. Although the authors of~\cite{kbim}
distance their work from full bimetric theories, in the limit of
perturbative fluctuations in a $K$-essence field the two perspectives
are equivalent. So it's natural to enquire which emergent metric theory
is associated with DBI models; and should we see such a metric as
the signature of a bimetric theory (as we shall do in the next Section),
which bimetric theory is associated with DBI models (or anti-DBI models; 
for the purpose of this discussion the sign of the coupling is irrelevant.)

There is a good reason why $K$-essence and DBI models are usually
only invoked when they reduce, rather than increase
the speed of sound: the innate fear of faster
than light propagation. Causality paradoxes, such as the
anti-telephone, are expected to arise (see~\cite{caus,kbim}; but
also see~\cite{recami} for an intrepid alternative). These can
be circumvented by re-interpreting $K$-essence  as a bimetric theory.
We review this mapping deriving the remarkable result that
the DBI model capable of producing scale-invariance, as shown in
Section~\ref{ksuccess} is mapped precisely into the minimal
bimetric theory discussed in Section~\ref{SBim}. However
{\it we stress that the mapping discussed in this Section is
perturbative, and an understanding of the full non-perturbative
aspects of the bimetric theory will only be achieved in the
next Section.}

The crux of the bimetric re-interpretation discussed
in~\cite{kbim} is the remark that the field equation of a
$K$-essence theory is equivalent to a Klein-Gordon theory subject
to effective metric: \be \label{tildeG}{\widetilde M}^{\mu\nu}=K_{,X} g^{\mu\nu} +
K_{,XX}\partial ^\mu\phi
\partial^\nu \phi\ee
with inverse
\be {\widetilde M}_{\mu\nu}=\frac{1}{K_{,X}}
{\left( g_{\mu\nu}-c_s^2 \frac{K_{,XX}}{K_{,X}}\partial_\mu\phi\; .
\partial_\nu\phi \right)} \ee
This metric is generally not conformal to the metric $g_{\mu\nu}$ servicing
gravity, immediately pointing us in the direction of bimetric VSL theories.
As long as \be 1+2X\frac{K_{,XX}}{K_{,X}} >0 \ee (equivalent to
requiring $c_s^2>0$)  the system is then hyperbolic with respect to
this metric and more generally the Cauchy problem is well defined,
resolving causality paradoxes. So it's tempting to identify 
(\ref{tildeG}) with  a conformal version of (\ref{gmnhat}),
with $\widetilde M_\mn$ playing the role of matter metric $\hat g_\mn$ and 
action (\ref{phikg2}).

We stress, however, that this mapping with bimetric theories is imperfect.
Firstly it induces a field equation
\be
{\widetilde M}^{\mu\nu}\nabla_\mu\nabla_\nu\phi=0
\ee
where the $\nabla$ operator is defined with respect to metric
$g_{\mu\nu}$, not $\widetilde M_\mn$. Secondly, a Klein-Gordon field
equation for $\phi$ doesn't correspond to a 
Klein-Gordon Lagrangian in bimetric theories,
as explained in Section~\ref{SBim}. Finally the subtleties in
defining the stress-energy tensor 
(see Section~\ref{SBim}) are altogether ignored.
Therefore the mapping just described can only be used gingerly.

A stronger connection is established in~\cite{kbim},
considering a 
background $\phi=\phi_0$ plus perturbations around it, a situation
well suited to cosmology. Leray's theorem states that the
perturbations satisfy 
\be {M}^{\mu\nu}D_\mu D_\nu \, \delta\phi=0 \ee 
with a new metric (conformal to ${\widetilde M}_\mn$) 
given by \be M^{\mu\nu}=\frac{c_s}{K_{,X}}{\left(g^{\mu\nu}+\frac{K_{,XX}}
{K_{,X}}\nabla^\mu\phi \nabla^\nu\phi\right)}\; , \ee where all
quantities are to be defined at
$\phi=\phi_0$. The covariant derivatives $D_\mn$ used in defining
the Klein-Gordon equation are now defined with respect to
metric $M_\mn$ and the theory may be derived from 
a Klein-Gordon Lagrangian fully built from $M_\mn$. 
The mapping is therefore rigorous and we should identify 
the inverse metric
\be\label{invMmn} M_{\mu\nu}=\frac{K_{,X}}{c_s}{\left(g_{\mu\nu}-c_s^2
\frac{K_{,XX}}{K_{,X}}\partial_\mu\phi \partial_\nu\phi\right)}
\ee
with the matter frame $\hat g_\mn$ via a linearization of (\ref{gmnhat}).
The action for the perturbations 
\be
\delta S^{(2)}=\int d^4 x{\sqrt {-M}}\frac{1}{2}
{M}^\mn\partial_\mu \delta\phi\, 
\partial_\nu \delta\phi
\ee
is then understood from a linearization of (\ref{phikg2}) or (\ref{phikg3})
via the proposed identifications
(we shall ignore the potential for the time being, as only
the kinetic term affects the mapping under study).

With this mapping in mind we can now ask 
onto what $K$-essence models one maps the minimal bimetric theory. 
In general {\it any} $K$-essence theory maps into a bimetric
theory of form (\ref{gmnhat})  but this could have a complicated form of 
$B=B(\phi,X)$. It is therefore interesting to ask which theories
lead to a minimal bimetric theory. If we compare (\ref{gmnhat}) with
(\ref{invMmn}), ignoring the conformal factor, we get:
\be - c_s^2
\frac{K_{,XX}}{K_{,X}}=-\frac{K_{,XX}}{{K_{,X}}+2 X K_{,XX}} =B \; .\ee
This is equivalent to 
\be \frac{K_{,XX}}{K_{,X}}=-\frac{B}{1+2BX} \ee and if $B$ is
constant this integrates 
immediately into: \be K=\frac{1}{B}\sqrt{1+2 B X} -\frac{1}{B} \ee
which is the DBI model (notice that the potential does not enter
the expression for the speed of sound and therefore is left
unconstrained).

This result has a remarkable implication. If we consider a minimal bimetric
theory where the bi-scalar is Klein-Gordon in the matter frame, then
in a perturbative scheme we should  identify ${\hat g}_{\mu}=M_{\mu\nu}$. 
Then we find that in the Einstein frame, $g_{\mu\nu}$, the theory
maps into a DBI model, i.e. precisely the $K$-essence model 
that leads to scale-invariance. Comparing with Section~\ref{ksuccess}
we see that we have $B=C$, so that a positive constant $B$ leads
to a DBI model with $f<0$, as required. This is a perturbative result,
but remarkable. It implies that scale-invariant fluctuations follow
from the minimal bimetric model, with a constant positive $B$, with
a minimal action defined in the matter frame.

As a side remark notice that had we taken the original
Clayton-Moffat theory, where the bi-scalar is Klein-Gordon in the
Einstein (or VSL) frame we would have discovered that in the matter
frame its dynamics maps into DBI model with $f>0$. This can be proved by a
straightforward adaptation of the above calculation, but the result
can be intuitively understood. In this original model the speed of sound
for the bi-scalar is the speed of gravity, which in the early universe
is much smaller than the speed of light. So the speed of sound
for the bi-scalar in the matter frame becomes negligible, 
just like for a DBI model with $f>0$.

\section{The full non-perturbative structure}\label{full}
We are now prepared to identify the full non-perturbative bimetric
structure behind the various models presented in this paper. It
will be simpler and more elegant, and perhaps we should have
presented it upfront, bypassing the hardships of cosmology.
We stress that the work in~\cite{kbim} is distinct from the bimetric
picture to be presented in this section, and that there should be
observational differences between the two, should there be further
matter fields, or non-perturbative effects be under study.

Let the dynamics of bi-scalar $\phi$ be driven by a pure
cosmological constant in the matter frame, that is: \be
S^1_\phi=\int d^4 x\sqrt{-\hat g}(-2\hat \Lambda)\; .\ee Even if
we don't bestow a kinetic term upon $\phi$ (either in the matter or
the Einstein frame) the field acquires dynamics, due to the subtlety
explained at
the end of Section~\ref{SBim}. In the absence of $K$ and $V$ 
the first two terms  in (\ref{modkg2}) disappear, but
since ${\hat T}^\mn=\Lambda{\hat g}^\mn$ the field satisfies
\be{\hat T}^\mn
{\hat\nabla}_\mu{\hat\nabla}_\nu\, \phi={\hat \Lambda}{\hat g}^\mn
{\hat\nabla}_\mu{\hat\nabla}_\nu\, \phi=0\; ,\ee
that is, a Klein-Gordon equation in the matter frame. This is
actually the only way to achieve a Klein-Gordon equation
of motion in the matter frame, since 
a Klein-Gordon Lagrangian, Eqn.~(\ref{phikg2}), does not 
translate into a Klein-Gordon equation.

In addition this theory has the required behaviour, for structure
formation, to all orders in
the Einstein frame. From (\ref{gmnhat}) we have \be {\hat
g}=g(1+2BX)\ee 
(see~\cite{bim,lang} for a derivation) 
so that $S_\phi^1$ in the Einstein frame becomes
\be S^1_\phi=\int d^4 x\sqrt{- g}\sqrt{1+2BX}(-2\hat \Lambda)\; .
\ee This is already all that is needed to obtain the correct high
energy behaviour and therefore scale-invariant fluctuations. If
additionally we want to ensure that the theory has the correct low
energy limit we must choose \be \hat \Lambda=-\frac{1}{2B}\ee and
require that an exactly balancing cosmological constant is present
in the Einstein frame. This results in the DBI action (\ref{dbiS})
with $f=-B$ (or $C=B$). As explained in Section~\ref{ksuccess}, we still 
need to add a mass term in the Einstein frame, with 
\be
m^2=\frac{3}{2BM_{Pl}^2}
\ee
but this is in fact required by the Bianchi identities, if we want
$\phi$ to remain dynamical in all circumstances.

To summarize all we need is a bimetric theory
with two balancing cosmological constants, one in each frame: \be
S_\phi=\int d^4 x\sqrt{-\hat g}\frac{1}{B} - \int d^4
x\sqrt{-g}\frac{1}{B}\; ,\ee 
tuned to the same constant $B$ which appears in the relation between
the two metrics (\ref{gmnhat}).
Such a theory has a Klein-Gordon dynamics
in the matter frame but becomes DBI in the Einstein frame. The
Bianchi identities then force it to have a mass term 
ensuring the appropriate background
scaling solutions at high energies. Its DBI behaviour in the
Einstein frame induces a varying speed of sound with the correct
profile to generate scale-invariant fluctuations with amplitude
\be \frac{2^{3/2}(5+3w)^2}{w(1+w)}\frac{1}{B M_{Pl}^4}\sim 10^{-10}
\ee (see Eq.~\ref{amp1}). 
Thus, with the possible exception of the equation of state $w$,
nothing remains undefined in this theory.

It is possible that the reinterpretation of the anti-DBI action 
in the bimetric framework might shed some light on its known 
pathologies~\cite{adams}
if interpreted in a single frame (the theory only becomes anti-DBI in the
gravity frame). This matter is beyond the scope of this paper.

\section{Discussion}

It is often stated that units can always be defined so that the speed of 
light is a constant. This is true in the same way that units can
always be defined so that the Hubble ``constant'' is indeed a constant,
the acceleration of gravity is the same everywhere as it is on Earth, 
the Universe is not expanding, etc, etc, etc. The argument is a perfect
twin and likewise the reason it makes sense to talk about varying $c$ is 
identical to why
it makes sense to talk about the expansion of the Universe, Newton's
laws, a varying $G$,  etc, etc, etc: they are dimensional statements 
justified by the simplicity of the overall picture. For example
by choosing units where the universe is not expanding all 
the equations of physics would become ridiculously complicated.

Nowhere is this more obvious,
with regards to VSL, than
in the present paper. We could always define units where
$c$ is constant: this is in fact the matter frame. But then the Einstein
equations become a royal mess, with the perturbation equations
looking even uglier; that the speed of gravitational
waves is now changing is an omen of what would then happen to the formalism. 
The gravitational dynamics of the theory (background and perturbations)
is crying out
for the VSL or Einstein frame to be used, just as it would 
if we recast modern cosmology into units in which the Universe
isn't expanding. This is far from a technicality.
For example it could be argued that the work in this paper is simply
inflation~\cite{infl} or ekpyrotic~\cite{ekp} 
in a different frame. This is obviously
nonsense: if we erase the VSL feature from the theory by going into
the matter frame Einstein's equations and all the perturbation
equations are horrendously modified. This ``dynamical'' feature is 
not part of any inflation or ekpyrotic model, and no doubt the 
discussion of scale-invariance would be rather different if they were 
to be included~\cite{mofstruc}.

We have used $K$-essence models, of the DBI persuasion, as a stepping
stone to the construction presented in this paper. But we believe that
the bimetric VSL structure identified towards the end of the paper offers
a superior description. To save $K$-essence from the perils of the causality
paradoxes the theory needs to be reinterpreted as an emergent~\cite{kbim} 
or a  bimetric theory.
The construction obtained is then valid only perturbatively and coincides
with ours to first order. But our bimetric construction is valid
to all orders, and in any case has a simpler dynamics: two balancing 
cosmological constants, one in the matter frame the other in the Einstein
frame, as discussed in Section~\ref{full}.

In this paper we focused on initial conditions set by a vacuum 
state which is at first inside the horizon. 
We leave to further work establishing the connection with thermal initial
conditions~\cite{csdot,pedro,param}.  This is because these require a phase 
transition in the speed of light, and are best realized in a 
theoretical framework diametrically opposed to the one described here
(see, for example,~\cite{hag,holo}).
The groundbreaking paper on VSL~\cite{jm} provides the best 
set up for implementing such a phase transition. In future work~\cite{ngr}  
we shall also present the predictions made by these models 
concerning non-Gaussianity and gravitational waves. 
After this paper was submitted new work appeared~\cite{Khpia}
where these issues are partly addressed. However the scope 
of~\cite{Khpia} is different from our paper; notably inflation is still
required.


I'd like to thank N. Afshordi,  R. Brandenberger, C. Contaldi, 
A. Tseytlin and A. Vikman for helpful comments. I'm very grateful to 
John Moffat for providing the inspiration for this paper and for
a careful reading of the manuscript.

\end{document}